\documentclass[aps,prl,twocolumn,superscriptaddress]{revtex4-1}
\usepackage{graphicx}
\usepackage{mathrsfs}
\usepackage{bm}
\usepackage{amsmath}
\usepackage{color}
\usepackage{dcolumn}
\usepackage{epstopdf}
\usepackage{dsfont}
\usepackage{amssymb}
\usepackage{tabularx}
\usepackage{array,ulem}
\usepackage{float}
\usepackage{colordvi}
\usepackage{verbatim}
\usepackage{hyperref}

\hypersetup{
colorlinks=true,
linkcolor=blue,
filecolor=blue,
urlcolor=blue,
citecolor=blue,
}

\begin{document}
\title{
Engineering Chiral-Induced Spin Selectivity in an Artificial Topological Quantum Well
}

\author{Lizhou Liu}
\affiliation{International Center for Quantum Materials, School of Physics, Peking University, Beijing 100871, China}

\author{Peng-Yi Liu}
\affiliation{International Center for Quantum Materials, School of Physics, Peking University, Beijing 100871, China}

\author{Tian-Yi Zhang}
\affiliation{International Center for Quantum Materials, School of Physics, Peking University, Beijing 100871, China}

\author{Qing-Feng Sun}
\email[Correspondence author:~~]{sunqf@pku.edu.cn}
\affiliation{International Center for Quantum Materials, School of Physics, Peking University, Beijing 100871, China}
\affiliation{Hefei National Laboratory, Hefei 230088, China}

\date{\today}

\begin{abstract}
Chiral-induced spin selectivity (CISS) is a striking phenomenon in which spin-unpolarized electrons become spin-polarized after traversing a chiral medium.
Theoretical studies have shown that spin-orbit coupling, geometric chirality, and dephasing act cooperatively for this effect to emerge.
Inspired by this, we demonstrate a solid-state realization of CISS in an engineered InAs/GaSb quantum well where geometric chirality and dephasing can be introduced controllably.
Introducing a chiral structure produces a clear spin polarization whose sign reverses when the chirality is flipped, and whose magnitude grows systematically with the number of dephasing electrodes, while achiral configurations exhibit no spin selectivity.
The polarization remains robust even under strong Anderson disorder, showing that the engineered chiral structures provides an intrinsically stable route to spin-selective transport.
These results establish a solid-state platform in the topological quantum well system for controllably generating the CISS effect.
\end{abstract}

\maketitle

\textit{Introduction---}Chiral structures play a central role across physics, chemistry, and biology, influencing electron transfer, molecular interactions, and reaction pathways~\cite{Hsieh2009, Bornscheuer2012}.
A striking manifestation of chirality is the chiral-induced spin selectivity (CISS) effect, in which spin-unpolarized electrons become spin-polarized when passing through a chiral medium without any magnetic fields or ferromagnetic elements~\cite{Ray1999Science, Gohler2011Science, Alam2015, Naaman2019NatRevChem,Xie2011NanoLett, Latawiec2025PNAS, Eckvahl2024JACS,Kettner2014JPCC, Mishra2013PNAS,Rana2025JACS, Jia2020ACSNano, Sun2024NatMater, Moharana2025SciAdv, Wang2025PRL,Albro2025ACSNano, BanerjeeGhosh2018, Eckvahl2023,Lu2019SciAdv,LiuTH1,LiuTH2}.
CISS has been observed across a wide range of chiral systems, including DNA-based structures~\cite{Gohler2011Science,Xie2011NanoLett, Latawiec2025PNAS, Eckvahl2024JACS},
peptides and protein-like biomolecules~\cite{Kettner2014JPCC, Mishra2013PNAS},
supramolecular assemblies and helical polymer networks~\cite{Rana2025JACS, Jia2020ACSNano, Sun2024NatMater, Moharana2025SciAdv, Wang2025PRL},
hybrid organic-inorganic interfaces~\cite{Albro2025ACSNano, BanerjeeGhosh2018, Eckvahl2023},
and halide perovskites~\cite{Lu2019SciAdv},
with the spin-polarization sign reversing upon chirality reversal.
These widespread experimental observations highlight the fundamental importance of chirality in governing spin-dependent electron transport, with implications for spintronics, enantioselective chemistry, and biological charge-transfer processes~\cite{Naaman2019NatRevChem}.

Over the past decade, various theories have been proposed to explain CISS,
including chiral-molecular spin-orbit coupling (SOC)~\cite{Guo2012PRL, Guo2014PNAS, Chen2023JPCL, Chen2024NanoLett, Liu2024CIP, Du2023CT}, interaction-driven mechanisms~\cite{Fransson2019JPCL}, phonon-assisted or polaron transport~\cite{Zhang2020Polaron, Das2022JPCC}, spin-flip scattering~\cite{Nurenberg2019PCCP}, and spinterface or interface effects at electrode–molecule junctions~\cite{Dubi1,Dubi2022ChemSci}, among others.
An effective theoretical framework suggests that the combined action of SOC, geometric chirality, and dephasing leads to the emergence of the CISS effect~\cite{Guo2012PRL, Guo2014PNAS}.
In this framework, SOC induces spin-dependent phases~\cite{addsun2005}, which, together with dephasing, result in asymmetric transmission of the two spin states~\cite{Guo2012PRL, Guo2014PNAS, Zhang2025JPCL}.
This framework quantitatively accounts for longitudinal CISS in DNA~\cite{Guo2012PRL, Guo2012b}, proteins~\cite{Guo2014PNAS}, and helicene molecules~\cite{Pan2015};
transverse CISS in current-in-plane devices and molecule-metal junctions~\cite{Wang2025PRL};
CISS-related spin-charge conversion phenomena ~\cite{Liu2025JPCL, Zhang2025ICISS};
and dynamical CISS in donor-bridge-acceptor systems and time-dependent spin-polarization dynamics~\cite{Zhang2025PRB, Liu2025}.
These successes demonstrate the broad applicability of the theory.
A natural question that follows is whether these insights can be used to engineer a fully controllable solid-state CISS device.

\begin{figure}
  \centering
  \includegraphics[width=7.2 cm,angle=0]{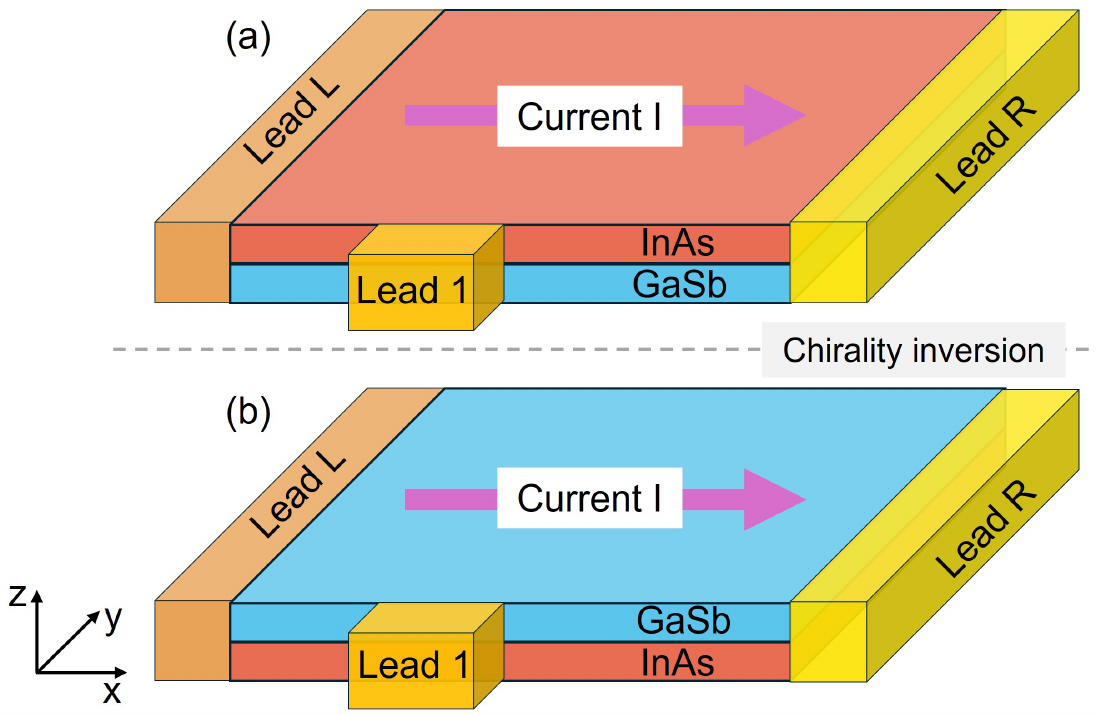}
  \caption{Schematic of two chiral configurations of the InAs/GaSb quantum well device used to demonstrate the CISS effect. (a) Left-chirality configuration: the InAs layer lies above GaSb, and the bottom dephasing electrode (Lead~1) is attached to the lower boundary.
 (b) Right-chirality configuration: the layer sequence is inverted, with GaSb on top of InAs, corresponding to an opposite (right-chirality) chirality. In both cases, the transport current flows from Lead L to Lead R.
}
  \label{fig1}
\end{figure}

In this Letter, we address this question by considering an InAs/GaSb quantum well engineered to possess structural chirality.
The inverted band structure of this system realizes a quantum spin Hall (QSH) phase, in which counterpropagating edge channels carry opposite spins due to spin-momentum locking and exhibit a quantized spin-Hall response.
This QSH phase has been firmly established in InAs/GaSb quantum wells through observations of quantized edge transport~\cite{Liu2008PRL, Knez2011PRL, Du2015PRL, Qu2015PRL}.
As illustrated in Fig.~\ref{fig1}, fixing the dephasing electrode (Lead 1) along one boundary breaks the mirror symmetry of the device, thereby introducing geometric chirality. The stacking order of the InAs and GaSb layers determines the chirality of the device.
We find that this engineered device exhibits clear CISS behavior: the chiral geometry induces strong spin polarization, its sign reverses when the chirality is flipped, and its magnitude increases with additional dephasing electrodes, whereas achiral configurations show no spin selectivity.
The effect remains robust under strong Anderson disorder, demonstrating that InAs/GaSb quantum wells offer a controllable solid-state realization of the CISS mechanism.
These findings show that CISS can be engineered and reversibly controlled in a solid-state quantum well, offering a practical route toward electrically tunable spin-selective devices.

\textit{Model and Hamiltonian---}
The chiral device geometry considered in this work is shown in Fig.~\ref{fig1}.
The left and right terminals fix the transport direction along $x$, and a dephasing electrode (Lead~1) is attached only to the lower boundary ($-y$ direction).
This asymmetric boundary coupling breaks the in-plane mirror symmetry $y\to -y$ and defines a geometric chirality in the transport ($x$--$y$) plane.
In addition, the InAs/GaSb stacking breaks the out-of-plane mirror $z\to -z$.
With these ingredients, the two configurations in Figs.~\ref{fig1}(a,b) form a pair of opposite chiralities, referred to as left-chirality and right-chirality devices, respectively.
It is worth emphasizing that the device is nonmagnetic and preserves time-reversal symmetry.

The QSH effect admits minimal lattice descriptions such as the Kane-Mele model~\cite{Kane2005,Kane2005a} and the Bernevig-Hughes-Zhang (BHZ) model for band-inverted quantum wells~\cite{Bernevig2006}.
Here, we adopt the BHZ model to describe the InAs/GaSb type-II band-inverted quantum well~\cite{Bernevig2006, Liu2008PRL, Chen2012, Chen2015}. For the left-chirality device, its Hamiltonian is:
\begin{equation}
H = \sum_{\bf i} c_{\bf i}^\dagger T_0 c_{\bf i}
+ \sum_{{\bf i}} \big( c_{{\bf i}+\hat{x}}^\dagger T_x c_{\bf i} + c_{{\bf i}+\hat{y}}^\dagger T_y c_{\bf i} + \mathrm{H.c.} \big),
\label{eq:TB}
\end{equation}
Here ${\bf i}$ labels a lattice site, and ${\bf i}+\hat{x}$ (${\bf i}+\hat{y}$) denotes its nearest neighbor along the $+x$ ($+y$) direction.
The onsite and nearest-neighbor hopping matrices are
\begin{align}
T_0 &= (C + 4D)\,\tau_0  \otimes \sigma_0 + (M + 4B)\,\tau_z  \otimes \sigma_0, \nonumber\\
T_x &= -D\,\tau_0 \otimes \sigma_0 - B\,\tau_z   \otimes \sigma_0 + \frac{iA}{2}\tau_x  \otimes \sigma_z, \nonumber\\
T_y &= -D\,\tau_0  \otimes \sigma_0 - B\,\tau_z  \otimes \sigma_0 - \frac{iA}{2}\tau_y  \otimes \sigma_0,
\end{align}
where $c_{\bf i}=(c_{E\uparrow},c_{H\uparrow},c_{E\downarrow},c_{H\downarrow})^T$
is the annihilation operator acting on the electron-like band and hole-like band.
Here $\tau_{x,y,z}$ and $\sigma_{x,y,z}$ denote Pauli matrices acting on orbital and spin degrees of freedom, respectively, while $\tau_{0}$ and $\sigma_{0}$ are the corresponding identity matrices.
For the right-chirality device, since it is related to the left-chirality one by mirror symmetry, one only needs to replace $T_x$ with $T_x^{\dagger}$ (or $T_y$ with $T_y^{\dagger}$).
We use the lattice BHZ parameters $A=0.28$, $B=1.0$, $C=D=0$, and $M=0.38$ from Ref.~\cite{Chen2015} (lattice units $a=1$).
Following Ref.~\cite{Chen2015}, if we set $a=1$ to correspond to $5\,\mathrm{nm}$ and the energy unit $E_0=26.4\,\mathrm{meV}$,
then the parameters in the continuous BHZ model in Ref.~\cite{Bernevig2006} are $B_{phys}=-Ba^2 \approx 660 \mathrm{meV\,nm^2}$, $A_{phys}=Aa \approx 37 \mathrm{meV\,nm}$, and $M_{phys}=-10 \mathrm{meV}$.
These parameters are consistent with those of the InAs/GaSb quantum well~\cite{Chen2015}.
In addition, all energies are measured in units of $E_0$, including
the disorder strength $W$ and the dephasing strength $\Gamma_d$ presented later.
As shown in Fig.~S1(a-c) of the Supplemental material (SM)~\cite{SM2025}, this parameter set lies in the QSH regime and helical edge states emerge~\cite{Liu2008PRL, Knez2011PRL, Du2015PRL, Qu2015PRL}.
In these edge channels, spin-momentum locking ensures that opposite spins counterpropagate along each boundary, yielding time-reversal-protected modes that dominate low-energy transport in the topological regime~\cite{Bernevig2006}.
Although bulk inversion asymmetry and structural inversion asymmetry terms may exist in InAs/GaSb, they do not alter the QSH phase \cite{Liu2008PRL}.

The multiterminal device consists of the central scattering region attached to left, right, and dephasing electrodes.
The left and right leads are modeled as semi-infinite ribbons with the same tight-binding form as the central region, allowing a slight parameter mismatch ($B_L=1.2$, $B_R=0.8$) to demonstrate generality. Their conducting nature in the energy window of interest is confirmed by the lead ribbon spectra in Fig.~S1 and Sec.~I of the SM~\cite{SM2025}.
Dephasing is introduced through Lead~1, which acts as a B\"uttiker voltage electrode: electrons entering this electrode are reinjected into the system, losing their phase and spin memory, while the net current through the electrode remains zero~\cite{Buttiker1986PRB, Buttiker1986PRL, Xing2008PRB}.

According to the Landauer-B\"uttiker formalism, the spin-resolved
current flowing into lead $p$ is
\begin{equation}
I_{p\sigma}
=\frac{e^{2}}{h}
\sum_{q} T_{pq}^{\sigma}(E)\,(V_{p}-V_{q}),
\qquad \sigma=\uparrow,\downarrow ,
\label{eq:LB_current}
\end{equation}
where $V_p$ is the electrochemical potential of lead $p$ and the total charge current is $I_p=I_{p\uparrow}+I_{p\downarrow}$.
The transmission coefficient from lead $q$ to lead $p$ with spin $\sigma$ is
$
T_{pq}^{\sigma}(E)
=\mathrm{Tr}\!\left[
\boldsymbol{\Gamma}_{p}(E)\,
G^{r}(E)\,
\boldsymbol{\Gamma}_{q}(E)\,
G^{a}(E)\,
\mathbf{P}_{\sigma}
\right],
$
and the total transmission is $T_{pq}(E)=\sum_{\sigma}T_{pq}^{\sigma}(E)$
\cite{Meir1992PRL, Fisher1981}.
Here $\mathbf{P}_{\sigma}$ is the projection operator onto spin $\sigma$.
For a nonmagnetic time-reversal-symmetric device in linear response, microreversibility yields the reciprocity relation $T_{p q}^\sigma(E)=T_{q p}^{\bar\sigma}(E)$ \cite{Buttiker1986, Jacquod2012}.
The retarded Green's function of the central scattering region reads
\begin{equation}
G^{r}(E)
=
\Bigl[
E\mathbf{I}
-\mathbf{H}
-\sum_{\alpha \in \{\mathrm{L},\mathrm{R},1\}}
\boldsymbol{\Sigma}_{\alpha}^{r}(E)
\Bigr]^{-1},
\label{eq:Gr_def}
\end{equation}
with $G^{a}(E)=[G^{r}(E)]^{\dagger}$.
Here $E$ is the incident electron energy (i.e. the Fermi energy), $\mathbf{I}$ is the identity matrix in the Hilbert space of the central scattering region, $\mathbf{H}$ is its tight-binding Hamiltonian, and $\boldsymbol{\Sigma}_{\alpha}^{r}(E)$ ($\alpha=\mathrm{L},\mathrm{R},1$) are the retarded self-energies describing the coupling to the left lead, right lead, and dephasing electrode, respectively.
The linewidth matrices describing the coupling to lead~$\alpha$ are
$
\boldsymbol{\Gamma}_{\alpha}(E)
=i\!\left[
\boldsymbol{\Sigma}_{\alpha}^{r}(E)
-\bigl(\boldsymbol{\Sigma}_{\alpha}^{r}(E)\bigr)^{\dagger}
\right].
$
The self-energies of the left and right electrodes,
$\boldsymbol{\Sigma}^{r}_{\mathrm{L(R)}}(E)$, are obtained from their surface
Green's functions using the recursive methods of
Refs.~\cite{Lee1981PRB}.
To model dephasing, we attach Büttiker virtual leads to lattice sites in the central region,
$
H_{d}=\sum_{{\bf i},k}\Bigl(\epsilon_{{\bf i}k} a_{{\bf i}k}^{\dagger}a_{{\bf i}k}+t_{d} a_{{\bf i}k}^{\dagger}c_{{\bf i}}+\mathrm{H.c.}\Bigr),
$
which is strictly spin independent and nonmagnetic.
Within the wide-band approximation, each contacted lattice site contributes the same energy-independent local self-energy,
$
\Sigma_{d}^{r}(E)=-\frac{i}{2}\Gamma_{d}\tau_0\otimes\sigma_0,
$
with $\Gamma_{d}=2\pi\rho_{d}\lvert t_{d}\rvert^{2}$~\cite{Xing2008PRB, Guo2012PRL}.

Since the dephasing electrode acts as a voltage electrode, its potential $V_1$ is determined by enforcing the current-conservation condition $I_1=0$ under the applied bias $V_{\mathrm{L}}-V_{\mathrm{R}}$ \cite{Xing2008PRB, Jiang2009a}.
In our setup, the left electrode serves as the source with a fixed voltage $V_{\mathrm{L}}=V$, while the right electrode is grounded with $V_{\mathrm{R}}=0$, so that the applied bias is $V_{\mathrm{L}}-V_{\mathrm{R}}=V$.
After eliminating $V_1$ using the voltage-electrode condition,
the spin-resolved conductances corresponding to the current flowing into the lead R are
\begin{equation}
G_{\sigma}(E)
=\frac{I_{\mathrm{R}}^\sigma}{V_{\mathrm{R}}-V_{\mathrm{L}}}
=\frac{e^{2}}{h}\left[
T_{\mathrm{RL}}^{\sigma}
+
T_{\mathrm{R}1}^{\sigma}\frac{T_{1\mathrm{L}}}{T_{1\mathrm{L}}+T_{1\mathrm{R}}}
\right].
\label{eq:G_sigma}
\end{equation}
The resulting spin polarization is defined as
$
P_{s}(E)=\frac{G_{\uparrow}(E)-G_{\downarrow}(E)}
              {G_{\uparrow}(E)+G_{\downarrow}(E)} .
$
From Eq.~(\ref{eq:G_sigma}), a finite $P_s$ originates from the electrode-mediated term involving Lead 1.
In the helical regime, spin-momentum locking ties the spin degree of freedom to the propagation direction of edge states on a given boundary.
Attaching Lead 1 to only one boundary in the helical regime makes the electrode-mediated contribution different for the two spin channels, which directly yields $G_{\uparrow}\neq G_{\downarrow}$ and hence a finite $P_s$.

%}

\begin{figure}
  \centering
  \includegraphics[width=8.5 cm]{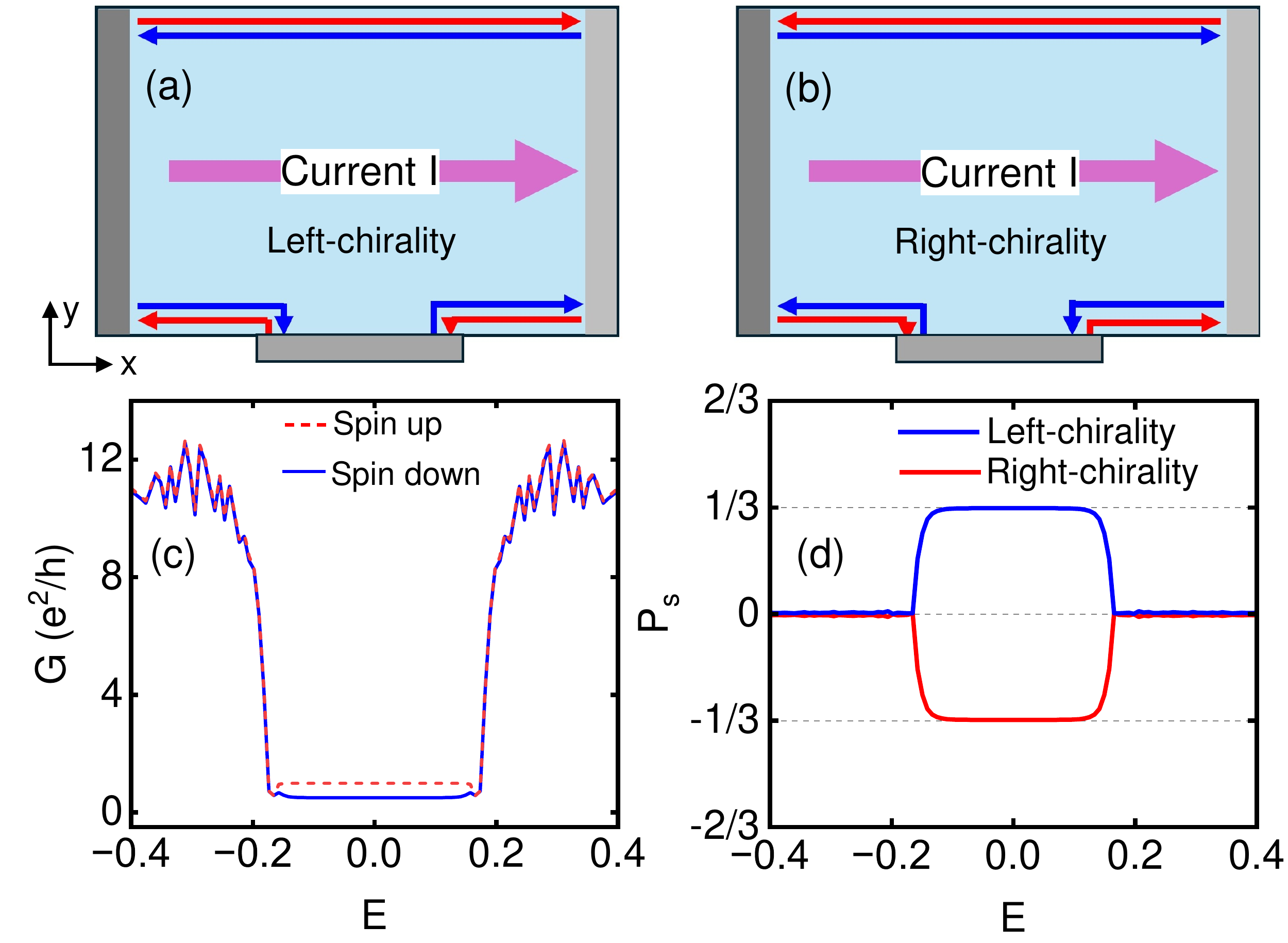}
  \caption{ (a,b) Helical-edge transport in the InAs/GaSb quantum well under the two opposite chiral configurations corresponding to Fig.~\ref{fig1}(a) (left chirality) and Fig.~\ref{fig1}(b) (right chirality).
  Spin-up (red) and spin-down (blue) channels propagate along opposite boundaries due to spin-momentum locking.
  (c) Spin-resolved conductances $G_\uparrow(E)$ and $G_\downarrow(E)$ for the left-chirality configuration under dephasing ($\Gamma_d = 0.5$), showing a clear splitting within the bulk-gap region.
  (d) Resulting spin polarization $P_s(E)$.
The central region contains 90 $\times$ 50 unit cells, and dephasing electrode covers 30 unit cells.}
  \label{fig2}
\end{figure}

\textit{CISS effect---}
To explore the emergence of the CISS effect in the topological InAs/GaSb quantum well,
we analyze the spin-dependent transport properties of the two chiral
configurations shown in Fig.~\ref{fig2}(a,b).
Lead~1 is attached to the lower boundary, and its fixed spatial position plays a central role in defining the geometric chirality of the device.
A clear spin asymmetry persists throughout the bulk gap: the helical channel propagating along the upper boundary retains nearly quantized transmission, while the channel propagating along the lower boundary is directly exposed to the influence of the dephasing electrode, leading to $G_{\uparrow}\neq G_{\downarrow}$ according to Eq.~(\ref{eq:G_sigma}).
In the left-chirality configuration (InAs above GaSb), the dephasing electrode couples to the forward-propagating spin-down edge state and the backward-propagating spin-up edge state.
In the right-chirality configuration, the reversed stacking order of InAs/GaSb relocates the helical edge states, causing the dephasing electrode to couple to the spin edge channels propagating in the opposite direction.
This results in opposite spin polarizations for the left- and right-chirality configurations.

Figure~\ref{fig2}(c) shows the spin-resolved conductances for the left-chirality case.
A clear splitting between $G_{\uparrow}(E)$ and $G_{\downarrow}(E)$ appears in the bulk-gap region, providing direct evidence of CISS induced by the chiral device geometry.
Outside the bulk gap, $G_{\uparrow}(E)$ and $G_{\downarrow}(E)$ also show differences, but the differences are minimal because the dephasing electrode, located at the edge, has a very weak influence on the bulk states.
The resulting spin polarization $P_s(E)$ [Fig.~\ref{fig2}(d)] attains sizable values near the midgap energy and remains almost unchanged over a broad energy window.
Reversing the stacking order changes the geometric chirality of the system, which leads to a corresponding sign reversal of $P_s(E)$.
This one-to-one correspondence shows that flipping the device chirality flips the polarization direction, a hallmark feature of the CISS effect~\cite{Naaman2019NatRevChem}.
In addition, our results are robust to the lead modeling, dephasing strength, and sample size (see Fig.S2-S5 and Secs.~II--IV of the SM~\cite{SM2025}).

\begin{figure}
  \centering
  \includegraphics[width=8.6 cm]{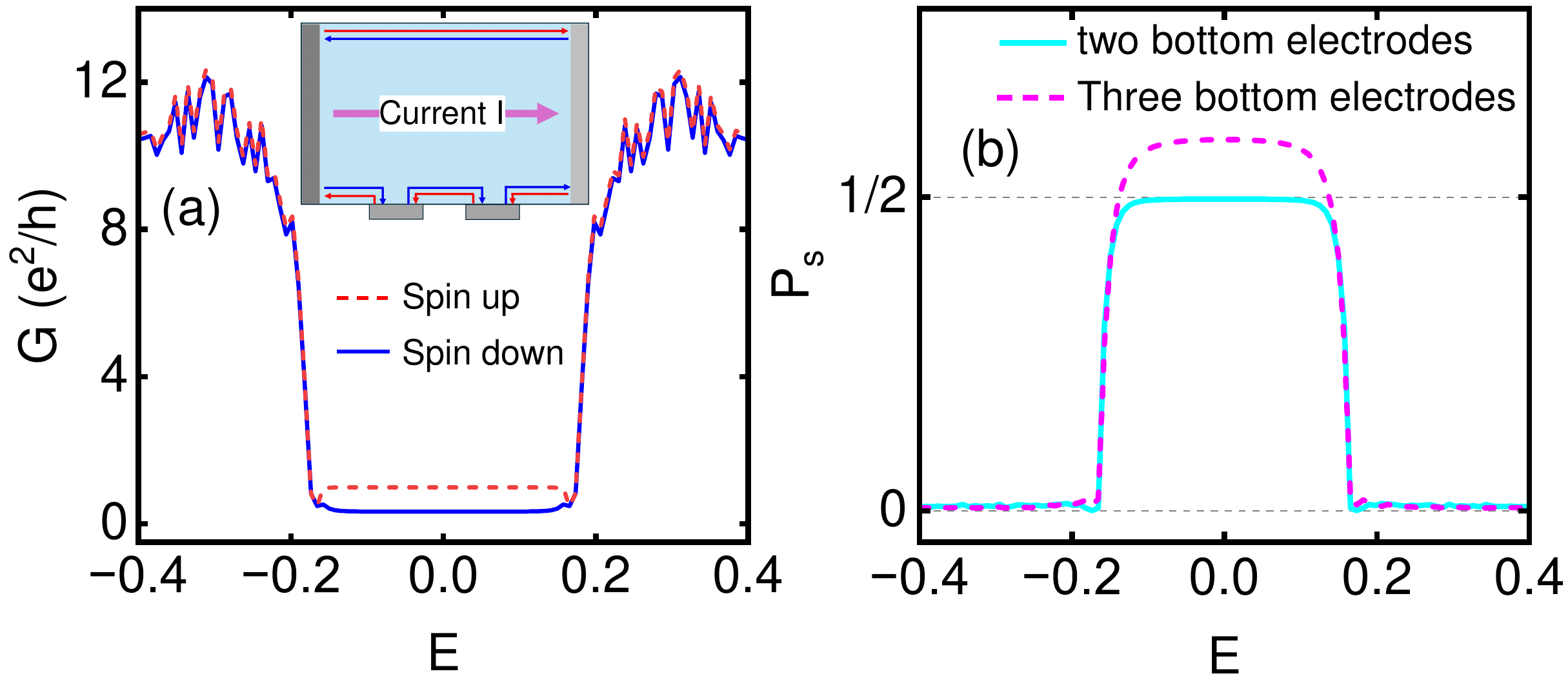}
  \caption{
(a) Spin-resolved conductances $G_{\uparrow}(E)$ and $G_{\downarrow}(E)$ for the left-chirality InAs/GaSb device with two bottom dephasing electrodes (inset) and $\Gamma_d = 0.5$.
  (b) Spin polarization $P_s(E)$ for devices with two (cyan) and three (magenta) bottom dephasing electrodes.
The central scattering region contains $150\times 50$ (two electrodes) or $210\times 50$ (three electrodes) unit cells, and each bottom dephasing electrode covers $30$ unit cells.}
  \label{fig3}
\end{figure}

\textit{Influence of the number of dephasing electrode---}
Next, we study how the number of dephasing electrodes placed along the lower boundary affects the spin-dependent transport.
Figure~\ref{fig3}(a) shows the spin-resolved conductances for a device with two bottom dephasing electrodes. A clear spin-dependent splitting appears inside the bulk-gap region.
To quantify the resulting change in spin selectivity, Fig.~\ref{fig3}(b) plots the corresponding spin polarization $P_s(E)$ for devices with two and three electrodes.
The polarization plateau increases with the number of electrodes, reaching about $P_s \approx 1/2$ for two electrodes and $P_s \approx 3/5$ for three electrodes (cyan and magenta curves).
These results demonstrate that increasing the number of dephasing electrodes enhances spin selectivity, as this is effectively equivalent to extending the length of the chiral structure \cite{Guo2012PRL, Guo2014PNAS}.
This trend reverses under chirality inversion, which flips the sign of $P_s(E)$.

The behavior in Fig.~\ref{fig3}(b) is captured quantitatively by a minimal multi-terminal Landauer-Büttiker model in which the bottom edge is replaced by a chain of $n$ voltage dephasing electrodes, while the top edge contains no dephasing electrodes.
The helical edge states yield perfect transmission with $T_{pq}=1$ between adjacent terminals.
Introducing $n$ dephasing electrodes along the lower edge imposes boundary condition $I_j=0$ ($j=1,2,\dots,n$) for each dephasing electrode, which fixes their voltages self-consistently and establishes a distributed potential drop along the lower boundary.
Solving the Landauer-Büttiker equations in Eq.~\eqref{eq:LB_current} yields a linear
potential profile $V_j=(n+1-j)/(n+1)$, so that the last electrode
adjacent to the lead R is at $V_{n}=1/(n+1)$.
The top helical branch contains no dephasing electrodes and therefore retains quantized transmission, while the bottom branch is attenuated successively by each electrode.
The resulting two-terminal conductances are
\begin{equation}
G_{\uparrow}=\frac{e^2}{h},\qquad
G_{\downarrow}=\frac{e^2}{h}\frac{1}{n+1},
\label{eq:Gupdown}
\end{equation}
leading to the spin polarization
\begin{equation}
P_s=(G_{\uparrow}-G_{\downarrow})/
         (G_{\uparrow}+G_{\downarrow})
      =n/({n+2}).
\end{equation}
This yields $P_s=1/3$, $1/2$, and $3/5$ for $n=1$, $2$, and $3$, respectively, in excellent agreement with the numerical plateaus in Figs.~\ref{fig2}(d) and \ref{fig3}(b).
Thus, the polarization enhancement arises from the cumulative dephasing applied to the helical branch selected by the chiral structure, with the effective strength increasing systematically as more dephasing electrodes are added.

\begin{figure}
  \centering
  \includegraphics[width=8.5 cm]{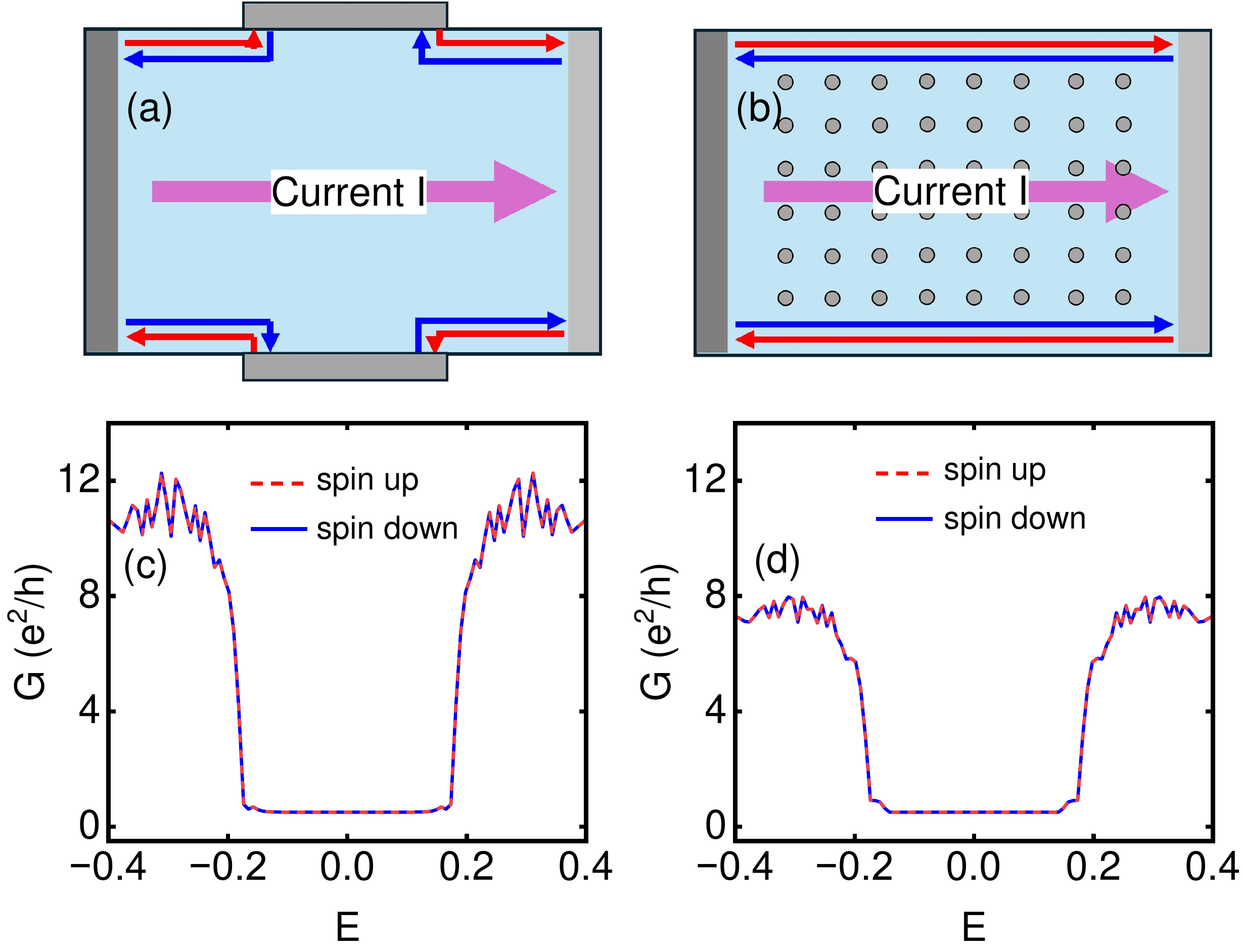}
  \caption{
  (a) Two-terminal device with dephasing electrodes attached symmetrically to the top and bottom edges.
  (b) Configuration in which dephasing electrodes are distributed uniformly within each $5 \times 5$ bulk plaquette.
  (c,d) Corresponding spin-resolved conductances $G_{\uparrow}$ and $G_{\downarrow}$ versus the Fermi energy $E$.
The other parameters are the same as in Fig.~\ref{fig2}.}
  \label{fig4}
\end{figure}

\textit{Achiral configuration---}
Next, we examine two achiral reference configurations shown in Fig.~\ref{fig4}, which allow us to isolate the contribution of geometric chirality to the spin polarization.
In the first configuration [Fig.~\ref{fig4}(a)], dephasing electrodes with identical coupling strength $\Gamma_d$ are attached symmetrically to both the top and bottom edges.
This symmetric arrangement removes any geometric chirality from the device.
This restores inversion symmetry between the counterpropagating helical channels, causing dephasing to act equally on both spin species.
As a result, the selective coupling responsible for the CISS response is eliminated, and the spin-resolved conductances $G_\uparrow(E)$ and $G_\downarrow(E)$ in Fig.~\ref{fig4}(c) become identical, yielding $P_s(E)=0$ across the entire energy range.

In the second configuration [Fig.~\ref{fig4}(b)], we introduce Büttiker virtual electrodes attached to lattice sites inside a $5\times5$ bulk region in order to emulate spatially distributed inelastic scattering.
Such bulk dephasing does not introduce any geometric chirality.
As a result, the conductance spectra in Fig.~\ref{fig4}(d) show no spin splitting, and the spin polarization remains essentially zero.

These results demonstrate that dephasing alone, whether applied along the edges or throughout the bulk, cannot generate spin selectivity.
A finite CISS response arises only when the device acquires geometric chirality, which makes the dephasing electrode couple to the edge channel on one boundary while the counterpropagating channel on the opposite boundary remains coherent.
Once this chiral asymmetry is removed, %both branches experience dephasing uniformly and
the spin polarization disappears immediately, confirming that geometric chirality and dephasing must act together for CISS to occur in the InAs/GaSb quantum well.

\begin{figure}
  \centering
  \includegraphics[width=8.5 cm]{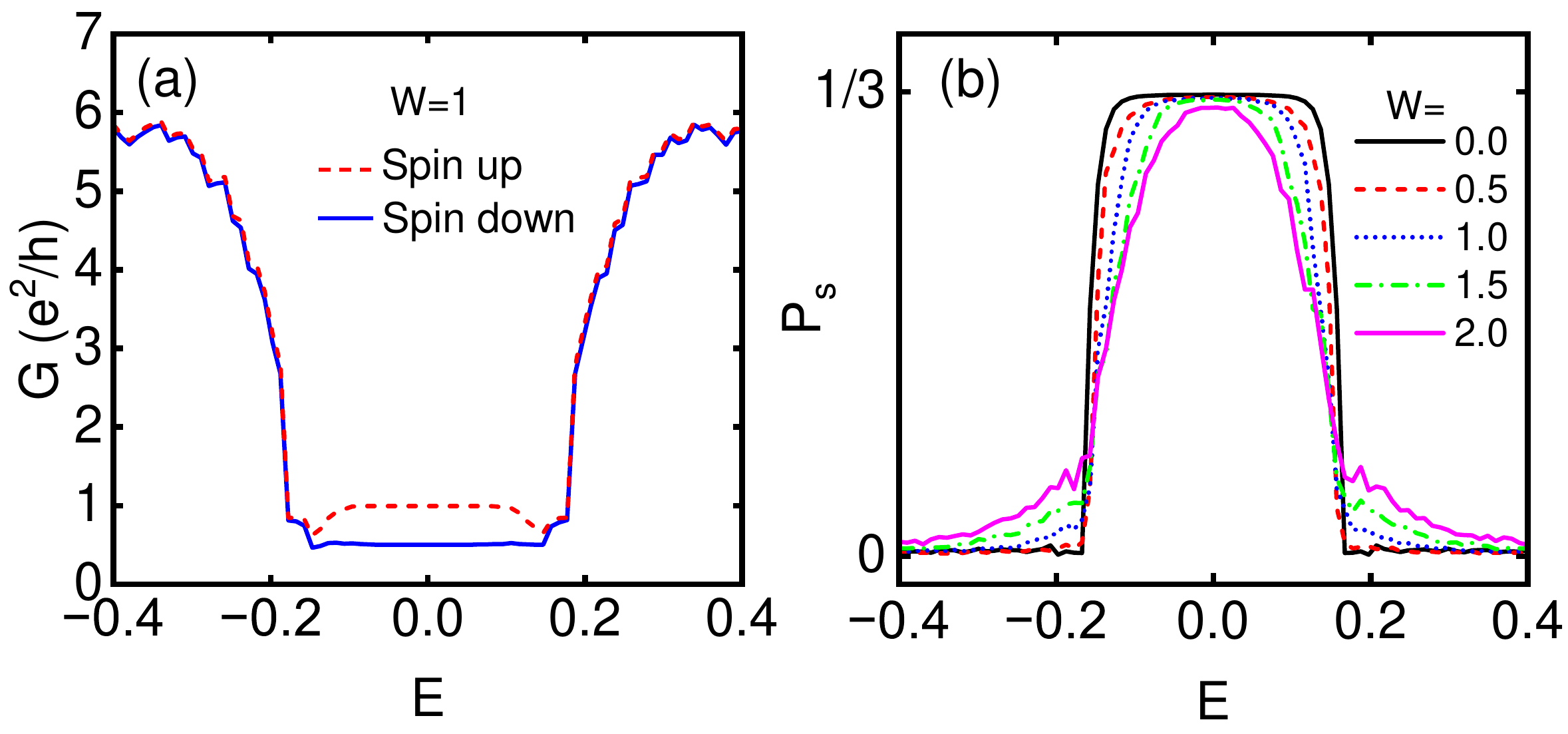}
  \caption{
  (a) Spin-resolved conductances $G_{\uparrow}(E)$ and $G_{\downarrow}(E)$ of the InAs/GaSb quantum-well device with a single dephasing electrode under a representative disorder strength $W=1$.
  (b) Spin polarization $P_s(E)$ versus the Fermi energy $E$ for different disorder strengths $W$.
 Both $\Gamma_d=0.5$ and $W$ are in units of energy $E_0$.
 The other parameters are the same as in Fig.~\ref{fig2}.}
  \label{fig5}
\end{figure}

\textit{Robustness against disorder---}
We next examine how Anderson-type disorder influences the CISS effect.
Random on-site potentials $W_i \in [-W/2,\, W/2]$ are added to the central region
$
H_{\mathrm{dis}} = H + \sum_i W_i\, c_i^\dagger c_i,
$
mimicking static potential fluctuations arising from impurities, interface roughness, or other structural imperfections.
Figure~\ref{fig5}(a) shows the spin-resolved conductances for a representative disorder strength $W=1$.
The overall spectral features remain similar to those of the clean system, and the splitting between $G_\uparrow(E)$ and $G_\downarrow(E)$ inside the bulk gap persists, indicating that the spin-selective response survives even under strong disorder.

The corresponding polarization $P_s(E)$ [Fig.~\ref{fig5}(b)] remains sizable and nearly unchanged for disorder strengths up to $W=2.0$.
Although minor fluctuations appear at larger $W$, the polarization plateau is largely preserved, demonstrating that the CISS response possesses strong robustness against disorder.
For dephasing $\Gamma_d$ that are not too weak, the disorder robustness depends only weakly on $\Gamma_d$ (see Sec.~V of the SM~\cite{SM2025}), consistent with time-reversal-protected helical edge transport~\cite{Bernevig2006,Hasan2010}.
As a result, the spin polarization in our engineered structure remains stable over a wide disorder range, reflecting the inherent resilience of helical-edge-mediated transport in InAs/GaSb quantum wells.
Such disorder tolerance is particularly advantageous for realistic device implementations, where potential fluctuations and interface imperfections are unavoidable.

\textit{Conclusion---}
In summary, we have theoretically demonstrated that an intentionally engineered chiral InAs/GaSb quantum well can serve as a solid-state platform for realizing the CISS effect.
The system integrates the three essential ingredients of SOC, chiral geometry, and dephasing, whose cooperative action generates robust spin-selective transport in the absence of magnetic fields or ferromagnetism.
By systematically varying the geometric chirality and the dephasing, we identified the microscopic conditions required for generating spin polarization in this topological system.
A finite and reversible spin polarization appears only in geometrically chiral configurations, while achiral structures exhibit no polarization.
The polarization magnitude increases with the number of boundary electrodes, consistent with analytical results from a multi-terminal Landauer--B\"uttiker model, and remains highly stable against disorder due to the protection of helical edge states.

These findings establish a direct link between CISS physics and mesoscopic topological transport.
They demonstrate that CISS can be intentionally engineered, tuned, and stabilized in nonmagnetic quantum wells through the controlled interplay of SOC, chiral geometry, and dephasing.
Our work provides a practical strategy for generating and manipulating spin-polarized currents in topological materials, offering a route toward chirality-based spintronic device architectures.

\textit{Acknowledgments---} This work was financially supported by the National Key R and D Program of China (Grant No. 2024YFA1409002), the National Natural Science Foundation of China (Grants No. 12374034 and No. 12547169), and the
Quantum Science and Technology-National Science and Technology Major Project (2021ZD0302403).
We also acknowledge the High-performance Computing Platform of Peking University for providing computational resources.

\end{document}